\documentclass[11pt,a4paper]{article}
\usepackage[utf8]{inputenc}
\usepackage{authblk}
\usepackage{graphicx} 
\usepackage[a4paper, total={7in, 10in}]{geometry}
\usepackage[english]{babel}
\usepackage{csquotes}

\usepackage{caption}
\usepackage{floatrow}   
\usepackage{subcaption}
\usepackage{graphicx,xcolor} 
\usepackage[framemethod=tikz]{mdframed}
\usepackage{amsmath}

\usepackage[natbib=true,style=numeric,sorting=none]{biblatex}
\def\e{\begin{equation}}
\def\f{\end{equation}}
\def\_#1{{\bf #1}}

\def\.{\cdot}

\usepackage{xcolor}
\usepackage{tabu}

\usepackage{siunitx}

\title{\large \textbf{Wide angle tolerant solar spectral splitter for lateral tandem solar cells}}
\author[1, *]{M. L. Schubert}
\author[1]{J. D. Fischbach}
\author[1]{M. Nyman}
\author[2]{L. L{\"u}er}
\author[2, 3]{C. J. Brabec}
\author[1, 4]{C. Rockstuhl}
\author[4, *]{T. J. Sturges}

\affil[1]{Karlsruhe Institute of Technology (KIT), Institute of Nanotechnology, Karlsruhe, Germany}
\affil[2]{Friedrich-Alexander-Universität Erlangen-N{\"u}rnberg (FAU), Institute of Materials for Electronics and Energy Technology (i-MEET), Erlangen, Germany}
\affil[3]{Helmholtz Institute Erlangen-N{\"u}rnberg (HI ERN), Erlangen, Germany}
\affil[4]{Karlsruhe Institute of Technology (KIT), Institute of Theoretical Solid State Physics, Karlsruhe, Germany}
\affil[*]{\textit{marie.schubert@kit.edu}; \textit{thomas.sturges@kit.edu}} 

\date{}

\begin{document}

\maketitle

\renewcommand{\abstractname}{}
\begin{abstract}

\noindent Maximizing the power conversion efficiency of solar cells plays a crucial role in upscaling solar energy production. Combining two or more solar cells with different bandgaps into a multi-junction tandem solar cells lowers thermalization losses and increases the power conversion efficiency. Whilst the best efficiencies have been achieved by vertically stacking solar cells, the fabrication process is technologically demanding and leads to high production costs. Novel photovoltaic materials such as organic photovoltaics allow solution processing, which enables the cost effective production of lateral multijunctions, where the single subcells are aligned side by side. To fully unlock their optimal performance, lateral tandems require careful light management, redirecting different spectral bands to the corresponding solar cell. So far, solar spectral splitters suffered from a strong angle dependency, which caused a degradation in performance at the slightest deviation from normal incidence. 
In this contribution, we reduce this limitation and achieve an enhancement in the conversion efficiency across a wide range of incident angles by inverse designing a solar spectral splitter comprised of two free-form microstructured surfaces on the top and bottom of a supporting glass substrate. Moreover, thanks to the versatility of our methodology, we can tailor the angle-dependent functionality of our device. As such, we also design devices that are optimized to provide enhanced performance at certain oblique angles, which correspond to different times of the day, e.g., when the unit price of energy is higher. 

\end{abstract}

\section{Introduction}\label{section: intro}
Single-junction solar cells are the current standard for solar energy production. However, their power conversion efficiency is limited to around 33\% by the detailed-balance limit for silicon cells \cite{Solar_energy}. This limit can be overcome by combining solar cells with different bandgaps into a multi-junction tandem solar cell. By using multiple subcells that are optimized for different portions of the total spectrum, these solar cells can absorb light across a broader range of wavelengths and with reduced thermalization losses \cite{lateral_tandem}. This multi-junction technique has provided the world record efficiency of 47.6\% for vertically stacked tandem solar cells under a light concentration of 665 suns \cite{ISE_worldrecord}. However, this type of tandem solar cell has the main disadvantage of being expensive and complicated to produce \cite{lateral_tandem}. The upper layer structure (including the electrodes) has particular requirements, as it needs to be transparent for some energies and absorbing for others. Current research on perovskite-silicon tandem solar cells solves some of these issues. Still, perovskite solar cells have challenges on their own, e.g., stability \cite{stabilityPerovskites} and the incorporation of toxic materials \cite{lead2020}.

Organic photovoltaics is an emerging photovoltaic technology approaching commercial viability as demonstrated by single cell efficiencies around $20\%$ \cite{20percentOranicSolarcell} and small module efficiencies of up to 14.46\% \cite{FAU_organic_worldrecord}. Recently, a "rainbow" tandem solar cell has been demonstrated, where two oraganic subcells were arranged laterally \cite{rainbow}. Lateral multijunction architectures are more cost-effective to produce than those stacked vertically, as they avoid the technological complexity of flawless multitude of layer deposition \cite{rainbow, lateralOrganicSC}. Thus, they are likely to be more competitive in the solar cell market \cite{lateral_tandem}. Furthermore, techniques like digital printing \cite{DigitalPrinting} could allow such lateral multijunctions to be patterned with micron resolution. 

Lateral tandems consist of two (or possibly more) subcells placed next to each other. However, because each subcell is optimized only for a particular portion of the solar spectrum, direct illumination by sunlight brings two key problems. Firstly, the subcell with the higher energy bandgap does not convert the lower energy portion of the spectrum at all. Secondly, the aforementioned thermalization losses continue to be an issue for the subcell with the lower energy bandgap. As such, without a light management strategy, lateral tandem solar cells cannot directly use the total solar spectrum, resulting in a much lower power conversion efficiency than vertically stacked ones.

To mitigate this drawback, a solar spectral splitter can be placed above the solar cells. Such a photonic component enhances the power conversion efficiency by redirecting different spectral components of the sunlight to different single-junction cells according to their absorption spectrum \cite{beamsplitter_experimental, BifacialSplitter2020, Lighttrapping2008, CompositeSplitter, dielectricmaterialsplitter}. Nowadays, powerful design methods exist to design structured interfaces that steer the incident light preferentially. These structured interfaces are diffractive and can be implemented either as a directly structured material layer \cite{KhoninaExploringDiffractive2024} or, alternatively, as a metasurface \cite{Yu2011, Khorasaninejad2016}.  

In a previous work, a single-layer solar spectral splitter was designed and measured to split sunlight into two spatially separated energy bands efficiently \cite{spectralsplitter}. However, the performance of this device rapidly degraded away from normal incidence. Essentially, oblique incident light continues to propagate into an oblique direction. In such a configuration, the spectrally redirected light hits an adjacent subcell, which is spectrally not optimized. Such a misrouting quickly causes the observed degradation.

In this contribution, we inverse design a two-layer solar spectral splitter on top of a lateral tandem solar cell. The parameters of the whole setup were chosen according to a theoretical study published recently \cite{lateral_tandem}. The spectral splitter enables the device to enhance its performance over a much wider range of angles. Conceptually, one can imagine that the first mask of the device redirects the incident light to a single-angle direction, for example, normal incidence. Next, the second mask spectrally distributes the light such that it can be harvested by the laterally arranged tandem solar cell. Notice that since we only optimize for the final functionality, the resulting device will not exactly incorporate this conceptual idea.

Topology optimization is an inverse design strategy used in the photonic community to find optical structures. Given a specific desired figure of merit (FOM), the inverse design problem aims to maximize or minimize the FOM of the system whilst obeying imposed constraints. The solution of the inverse design problem leaves us with a locally optimal structure of the setup \cite{inverse_design}. Specifically, we calculate the optical fields with Fourier optics and the thin element approximation, and then use gradient-based algorithms to iterate towards an optimized device. All the gradients are obtained with one forward and one backward (so-called `adjoint') pass through the simulation \cite{inverse_design2}. This simulation technique and gradient calculation enable us to optimize over many wavelengths and incident angles on a timescale of minutes.

When optimizing a device to maximize its performance at normal incidence $\pm 1^{\circ}$, we achieve a current gain of $69.40\%$ compared to a device without a solar spectral splitter, as long as the incidence angle does not exceed the optimization limits. However, a device that only performs at normal incidence is only useful when mounting the device on a motor-controlled stage to track the motion of the sun. Instead, we focus on optimizing a device that performs over a wide range of incident angles. In our case, we show a boost in current within $\pm 20^{\circ}$ of $47.69\%$.
In addition, we exploit the flexibility of our methodology to design devices that preferentially boost the performance at particular angles, which could correspond to times of the day when energy prices are generally higher.

\section{Results and Discussion}\label{section: results}

\begin{figure} [ht]
  \centering
  \begin{minipage}[t]{.6\linewidth}
      {\includegraphics[width=1.0\linewidth]{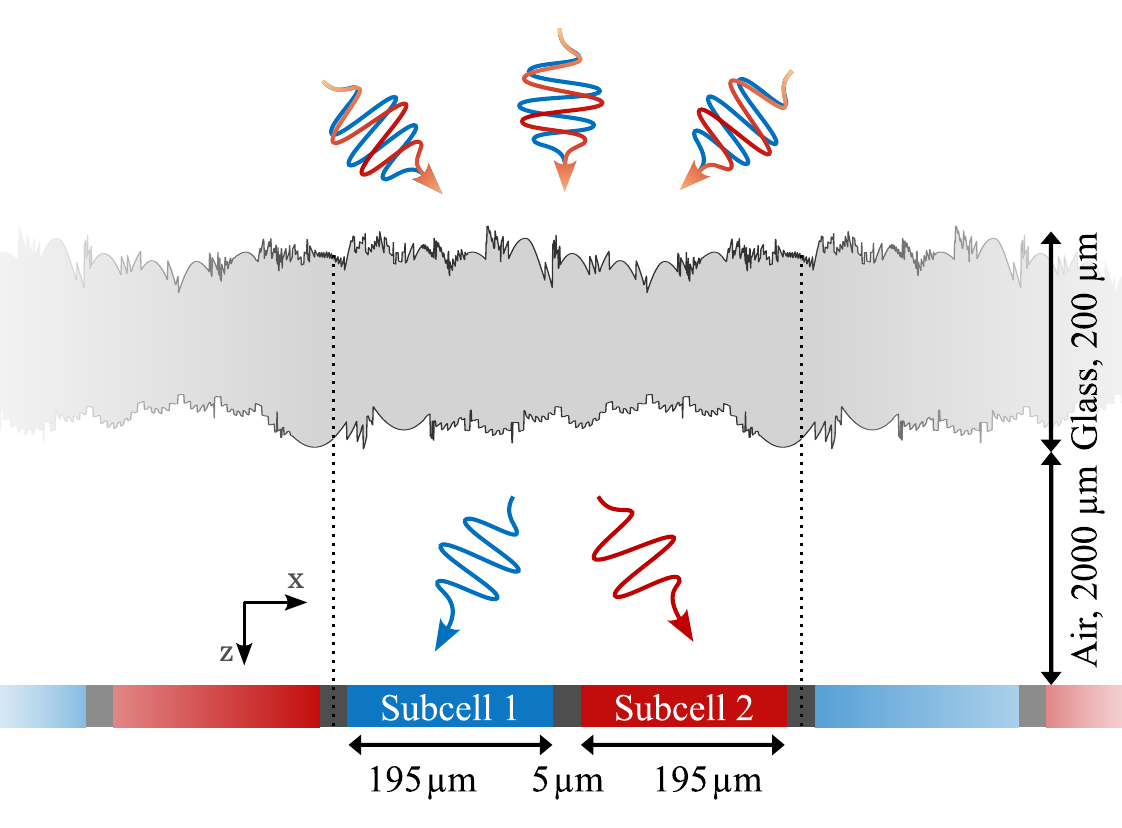}}%
  \end{minipage} 
  \hfill
  \caption
    {Schematic of the considered setup with the inverse designed solar spectral splitter and the lateral solar cells. The blue subcell 1 absorbs at shorter wavelengths, and the red subcell 2 absorbs at longer wavelengths.
    }%
  \label{fig:figure1}%
\end{figure}

We aim to design a solar spectral splitter with high spectral splitting efficiency across a wide range of incident angles. When combined with lateral tandem solar cells, this will result in a device that provides a higher power conversion efficiency over various relative positions of the sun. Figure~\ref{fig:figure1} shows a schematic of our device and a conceptual overview of the research task. The solar spectral splitter consists of two phase masks placed sequentially along the optical path, implemented by free-form structured surfaces at the top and bottom layer of a \qty{200}{\micro\meter} long \textrm{SiO$_{2}$} substrate. As we will see, having more than one phase-modulating surface is essential for achieving a wide-angle tolerance. Two tandem solar cells are placed directly underneath the spectral splitter at a distance of \qty{2000}{\micro\meter}. The solar cells are \qty{195}{\micro\meter} wide with a dead zone at each end of \qty{2.5}{\micro\meter}. \textcolor{black}{The two subcells form a multi-junction architecture and are connected in series. Organic semiconductor materials encompass a broad range of optical bandgaps and energy levels. Therefore, we have a large design freedom for experiments. Here, we assume an ideal case, where} the solar cells convert wavelengths solely within their operational range, specifically \qty{310}{\nano\meter} to \qty{780}{\nano\meter} for the left (blue) subcell, and \qty{780}{\nano\meter} to \qty{1340}{\nano\meter} for the right (red) device. Therefore, the ideal splitter would redirect all the short wavelength components of the illumination to the left subcell and all the long wavelength components to the right subcell. This functionality should be provided, ideally, independent of the incident angle of the light. Note that the structure shown in Fig.~\ref{fig:figure1} is one unit cell within a spatially periodic device. Finally, we emphasize that all these parameters are only exemplary. Although the quantitative results depend on these parameters, our design framework applies to any other related device.

Specifically, for a given light spectrum, our goal is to maximize the average current generated by the tandem solar cells across the target angle range. The spectrum that we consider is the typical AM 1.5 spectrum of the sun \cite{Solar_energy}, weighted by a factor $\sin(\theta)$ to account for the total reduction in power due to the tilt $\theta$ of the device away from normal incidence. We sample this spectrum at $N_\lambda = 100$ equidistant wavelengths between \qty{310}{\nano\meter} and \qty{1340}{\nano\meter}. We initialize the incoming field for a given incidence angle as a sum of plane waves with amplitudes corresponding to the weighted AM 1.5 spectrum. We then simulate the propagation of this field through the system, using Fourier optics and the thin element approximation, and record the wavelength-dependent field intensity arriving at each subcell, see Section \ref{section: fourier and thin element}. From this, we can calculate the current generated by the tandem solar cells by integrating the radiant flux over all wavelengths, see Section \ref{section: solar cell current}. Finally, we repeat this process for the $N_\theta$ angles and calculate the current averaged over the target angle range. This current constitutes our figure of merit, which we aim to maximize by using inverse design techniques to optimize the surface structures of the spectral splitter.

To enable efficient optimization, our code is developed using JAX \cite{jax2018github}, a software library that automatically differentiates native Python and NumPy code. The differentiable formulation allows us to compute the derivatives of the simulation results with respect to the height profiles of the two phase masks \cite{inverse_design, inverse_design2}. By performing a single ``backward" pass through the simulation code, we can obtain all derivatives simultaneously, maintaining computational complexity similar to that of the original function, regardless of the number of design variables used to parameterize the phase mask \cite{Griewank2008, Baydin2018}. As such, we use a gradient descent algorithm to iteratively refine the phase masks to a design that maximizes the total current.

To motivate our research goal and emphasize the necessity of two phase masks, let us start by presenting results for a single phase mask and highlight the problems that arise when considering oblique angles of incidence. Specifically, we optimize only the top layer of the spectral splitter whilst leaving the bottom layer as a flat interface, thus imparting no phase shift. 

\begin{figure}
  \centering
  \begin{minipage}[t]{0.99\linewidth}
      {\includegraphics[width=1.0\linewidth]{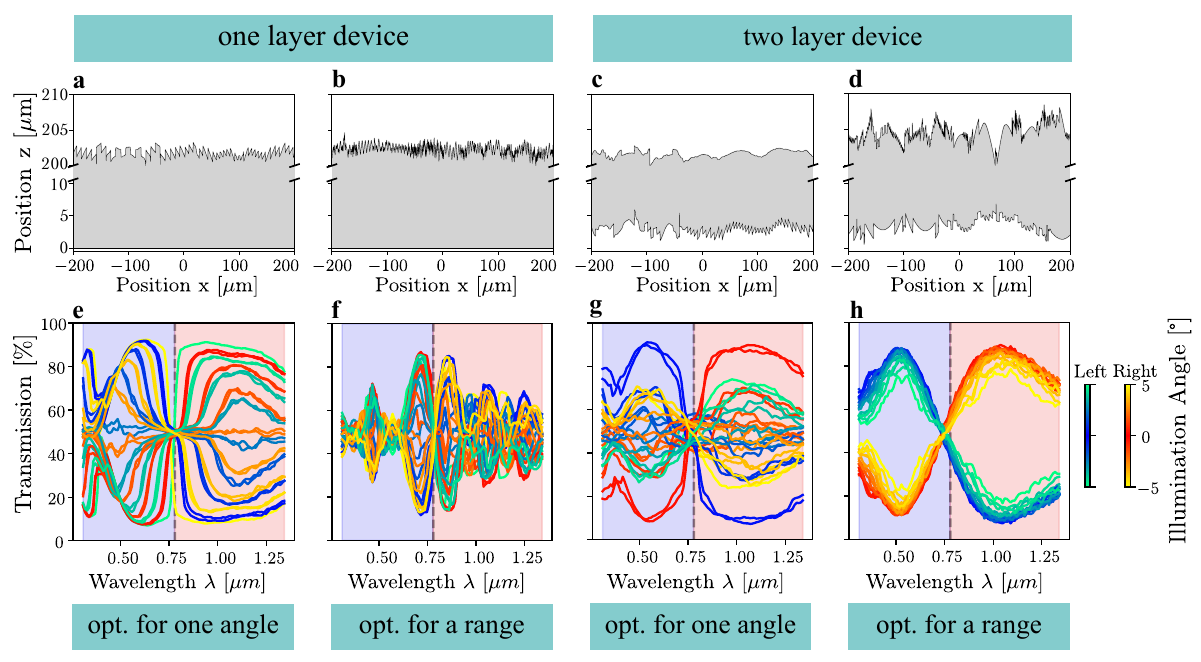}}%
  \end{minipage} 
  \hfill
  \caption
    {Comparison of different devices. The top row shows the device layout. \textcolor{black}{The corresponding maximum peak-to-valley height can be found in Table~1 of the Supplementary Information.} The bottom row shows the performance. (a) One-sided phase mask optimized for $0^{\circ}$ only. (b) One-sided phase mask optimized for 50 angles between $-5^{\circ}$ and $5^{\circ}$. (c) Two-sided phase mask optimized for $0^{\circ}$ only. (d) Two-sided phase mask optimized for 50 angles between $-5^{\circ}$ and $5^{\circ}$. (e) Performance of the one-sided solar spectral splitter optimized at an angle of $0^{\circ}$ only. (f) Performance of the one-sided solar spectral splitter optimized for angles between $-5^{\circ}$ and $5^{\circ}$. (g) Performance of the two-sided solar spectral splitter optimized at an angle of $0^{\circ}$ only. (h) Performance of the two-sided solar spectral splitter optimized for angles between $-5^{\circ}$ and $5^{\circ}$. 
In (e)--(h), the transmission onto the left (blue and aqua curves) and right (red and orange curves) subcell is shown separately for a total of 15 incident fields illuminating the device at equidistant angles between $-5^{\circ}$ and $5^{\circ}$. These are not identical with the optimization angles. }%
  \label{fig:figure2}%
\end{figure}

Firstly, we design this structure to provide an optimal performance at normal incidence only, without considering its performance at any oblique angles. The resulting design is shown in Fig.~\ref{fig:figure2}a. \textcolor{black}{It mostly consists of several regions of blazed-grating-like sawtooth structures reminiscent of two Fresnel lenses. The difference between the minimal and maximal height of the structure is \qty{3.24}{\micro\meter} along the z axis. We call this quantity maximum peak-to-valley height.} Fig.~\ref{fig:figure2}e shows the performance of this design when optimizing for 15 different incident angles equidistantly spread between $-5^{\circ}$ and $5^{\circ}$. To ensure that the optimization does not only work for the optimized angles, the testing angles shown in the figure are chosen to be different. The angles of illumination are indicated by the different color schemes from yellow to red back to yellow and from aqua to blue to aqua. The blue curves illustrate the transmitted light onto the left solar cell 1, while the red curves represent the light transmitted onto the right solar cell 2. An ideal device would correspond to a transmission plot for each subcell that looks like a step function, regardless of the incidence angle. For the left subcell (indicated by the blue background shading), this would correspond to 100\% transmission for the short wavelengths and 0\% transmission for the long wavelengths. We find that this device performs very well at normal incidence (shown by the dark blue and red curves), in agreement with the results from Ref.~\cite{spectralsplitter}. However, as we see in Fig.~\ref{fig:figure2}e, there is a rapid degradation in the performance away from normal incidence. As we approach angles of $\pm 2.5^{\circ}$ we do not observe any splitting behavior any more. Indeed, the results further deteriorates for angles around $\pm5^{\circ}$ (shown by the aqua and orange curves), where we see that the splitting is in fact totally inverted. 

To combat this issue, we try to explicitly account for the device's performance away from normal incidence by encoding it into the objective function. Specifically, we calculate the current generated for a range of different illumination angles and then consider our objective function as the average current across this range of angles. Such a device optimized for $50$ angles in the range $\pm5^{\circ}$ is shown in Fig.~\ref{fig:figure2}b. \textcolor{black}{The resulting device has a very rough, non-periodic profile with a maximum peak-to-valley height of \qty{4.62}{\micro\meter}.} Unfortunately, as we see in Fig.~\ref{fig:figure2}f, this approach is unsuccessful for a single phase mask. The optimizer cannot find any suitable device and instead delivers a device that has an entirely random and unhelpful performance at any incidence angle in the range of $\pm5^{\circ}$. Indeed, we conjecture that it may be simply impossible to achieve the desired functionality with a single freeform-surface phase mask.

This realization motivates us to study two-layer phase masks in the rest of this article. To motivate the second key ingredient in our final design strategy, we present results for a two-layer phase mask optimized only for normal incidence. The design and results are shown in Fig.~\ref{fig:figure2}c and ~\ref{fig:figure2}g. \textcolor{black}{The top layer design is much smoother than the bottom one. The maximum peak-to-valley height is \qty{2.74}{\micro\meter} for the top and \qty{4.73}{\micro\meter} for the bottom layer.} At $0^\circ$, we observe a distinct splitting of the different wavelengths onto the different solar cells. It looks like an ideal boxcar result. However, the results quickly become much worse for larger illumination angles, at some point resulting in a (noisy) flat response, corresponding to a zero splitting efficiency. At even larger angles around $\pm5^\circ$, the transmission response is again inverted, but much less so compared to the single-layer design. In conclusion, even when only optimized for normal incidence, the two-layer device already performs better than the one-layer device, both in terms of its splitting efficiency at normal incidence, and in terms of its angle insensitivity. Nonetheless, it is still not particularly robust towards different illumination angles. This is not surprising considering we did not optimize it to be so.

Therefore, in the same way as we attempted for a single phase mask, we can optimize a two-layer device that maximizes the current averaged over a range of angles. Again, we consider $50$ angles in a range between $\pm5^{\circ}$. The design and results are shown in Fig.~\ref{fig:figure2}d and ~\ref{fig:figure2}h. \textcolor{black}{The design exhibits similar roughness on both layers. The maximum peak-to-valley height is \qty{8.47}{\micro\meter} for the upper and \qty{6.23}{\micro\meter} for the lower layer.} We observe the desired splitting of the different wavelengths onto the different subcells, with consistent behaviour across the entire range of angles. The plot in Fig.~\ref{fig:figure2}h fulfills our expectation, with maximal transmission for the blue curve in the low wavelength range and maximal transmission for the red curve in the longer wavelengths. Naturally, the performance is better for those angles nearer the central optimization angle (normal incidence in this case). However, it is encouraging to note that a substantial increase in current over a device without solar spectral splitter is maintained across the entire range of angles, as we will show below. Compared with the device optimized for one angle (Fig.~\ref{fig:figure2}c,g), this device optimized for a range of angles (Fig.~\ref{fig:figure2}d,h) has a less boxcar-like shape at normal incidence, corresponding to a smaller spectral splitting efficiency. However this performance is now consistently maintained across the entire angle range $\pm 5^\circ$.

Having established and motivated our design strategy, we now investigate the robustness of this approach for even larger ranges of incidence angles. The idea is to show how our optimization potentially works for real-world applications in lateral tandem devices, where an enhanced performance is desired over wide angle ranges without the need for solar tracking. Therefore, we study the relative current gain, defined as the difference between the current provided by the device with an optimized spectral splitter and the current from the baseline, divided by the current from the baseline device which employs no spectral splitter. 

\begin{figure} [ht]
  \centering
  \begin{minipage}[t]{.6\linewidth}
      {\includegraphics[width=1.0\linewidth]{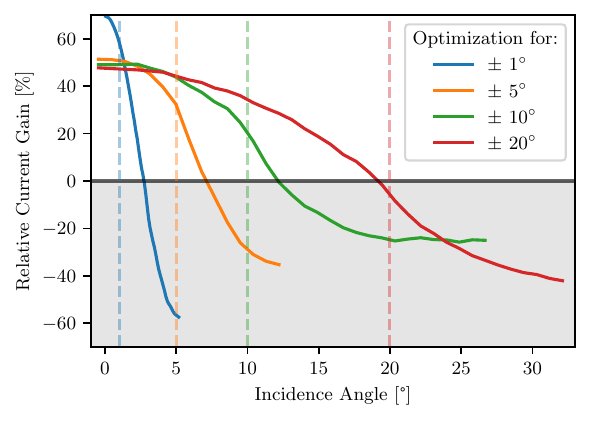}}%
  \end{minipage} 
  \hfill
  \caption
    {Relative current gain as a function of the angle of incidence for a set of devices optimized over different angle ranges. The dashed lines indicate the optimized angle range. The grey horizontal line and area indicate where no relative current gain is achieved.}%
  \label{fig:figure3}%
\end{figure}

Figure~\ref{fig:figure3} shows the relative current gain as a function of the angle of incidence for a few different devices optimized to perform over different ranges of angles, from narrow to wide. The dashed vertical lines indicate the width of the angular range considered in the optimization. The vertical grey line represents exactly $0\%$ of relative gain. All the points on the curves below this grey line correspond to a performance that, for such an oblique angle, is worse than having no spectral splitter at all. 

Most devices achieve a relative current gain within their optimization ranges, but their gain quickly degrades outside this window. The trade-off between a peak performance for narrow optimization ranges, versus a sustained performance across wider angles is clearly demonstrated in Fig.~\ref{fig:figure3}. For a small optimization range of $\pm 1^{\circ}$, the relative current gain is around $70\%$ for incidence angles of $\pm1^{\circ}$, but this performance immediately decays for larger incidence angles. For larger optimization ranges, we only achieve a gain of ca. $50\%$ for illumination of $\pm1^{\circ}$, but the decay of gain does not immediately occur. This means the more constrained the angular range considered in the optimization, the better the performance at the targeted angles. The wider the angular range in the optimization, the less pronounced the peak performance, but the longer the performance is maintained. On the other hand, it remains encouraging to note that when increasing the angular range from $\pm5^{\circ}$ onwards, the degradation of the peak performance is not well pronounced anymore, and it lowers only marginally. That being said, we see that for the largest possible angle range, it is not possible to see a positive relative current gain in the entire angular domain considered for the inverse design. This could be a hint that the constraints we impose in the optimization are too severe. A larger number of structured interfaces might mitigate the problem. However, this would come with a general device complexity that is undesirable. At the same time, due to the conservation of optical etendue, it is not possible to optimize the spectral splitter for the full, desired spectrum of $\pm90^{\circ}$ for the setting given in this contribution \cite{Entendue_PETERS}.

To further highlight the possible use of such devices for real-world applications, and the flexibility of our design method, we optimize devices that have a peak performance at an oblique angle of incidence. Depending on the orientation of the device, we might envisage that such angles are suitable for devices geared to maximize the power conversion in the morning or evening hours when the cost of energy is typically higher \cite{market_price}, and an increase in current production could be especially beneficial. Furthermore, this could be useful for building integrated photovoltaics, where modules will not necessarily have normal incidence at peak sun illumination.

\begin{figure} [ht]
  \centering
  \begin{minipage}[t]{.6\linewidth}
      {\includegraphics[width=1.0\linewidth]{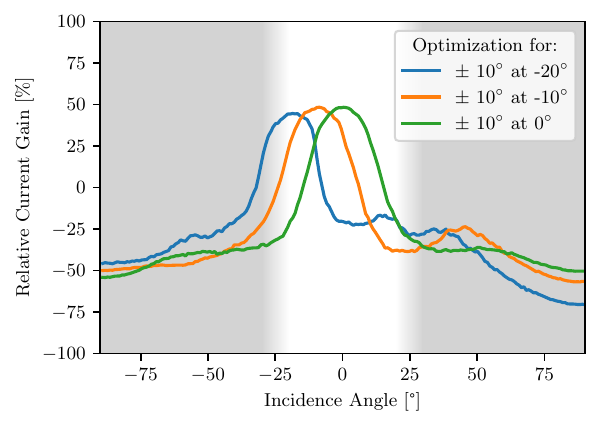}}%
  \end{minipage} 
  \hfill
  \caption
    {Optimization of solar spectral splitters with target incidence angles $\pm 10^{\circ}$ centered around $-20^{\circ}$, $-10^{\circ}$ and normal incidence $0^{\circ}$. The grey color shading starting at $\pm 20^{\circ}$ towards larger angles shows the regime where the thin element approximation becomes increasingly less reliable.
    }%
  \label{fig:figure4}%
\end{figure}

Figure~\ref{fig:figure4} depicts the relative current gain for devices optimized over an angle range of $\pm 10^{\circ}$ centered at $0^\circ$, $10^\circ$, and $20^\circ$ degrees. Note the grey shading starting at around $\pm 20^{\circ}$, where the thin element method becomes less reliable, see Section~\ref{section: fourier and thin element}.
Each device shows a relative current gain for its corresponding optimized angle range. Thus, our method works for ranges of incident angles that do not need to be centered around normal incidence. However, a slightly higher and broader gain can be noted for the optimization around normal incidence. This can be explained by the fact that more slanted angles are difficult for the setup to optimize since they need stronger rerouting by the solar spectral splitter.

In conclusion, we used inverse design techniques to design wide-angle tolerant solar spectral splitters that enhance the power conversion efficiency over a large range of solar  radiation incidence angles. In particular, we demonstrated a device that enhances the current-generation performance for nearly $\pm 20^\circ$. We expect that this range can be pushed even further. However, we refrained from analyzing such cases as the thin element approximation we utilized in our design method starts to become unreliable beyond $\pm 20^\circ$. Depending on the requirements of any given device, one can tailor the width of the angle range over which an optimized device provides a performance enhancement. We showed that the difference in peak performance between, e.g., $\pm 5^\circ$ and $\pm 20^\circ$ was quite minimal, and thus, one can obtain good performance for even very wide angle ranges. Lastly, we demonstrated the flexibility of our design method, with the ability to optimize devices that have a peak performance at oblique angles of incidence, perhaps corresponding to particular times of the day when the unit energy cost is highest. We highlighted that two phase masks are necessary for enabling the wide-angle tolerance of a solar spectral splitter for lateral tandem solar cells, combined with a design strategy that explicitly optimizes for a range of angles rather than just normal incidence. At this moment, all our optimizations are performed in 2D. We expect our setup to easily expand to the three-dimensional case, where, for example, both the altitude and azimuth angle of the sun can be considered. This would take our devices even closer to real-world applications.

\section{Methods}\label{section: methods}
\subsection{Fourier optics and thin element approximation}\label{section: fourier and thin element}
Within scalar diffraction theory, Fourier optics is a mathematical tool describing the propagation of light in the spatial frequency domain \cite{fourier_optics}. Let us consider a monochromatic field $E_0(x,\lambda)$ that has the plane wave decomposition (obtained through the Fourier transform) $\tilde{E}_0(k_x, \lambda)$ with $k_x$ the wavenumber and $\lambda$ the wavelength. This field, along with the structures we consider, is invariant in the out-of-plane $y$-direction. The propagation of $\tilde{E}_0(k_x, \lambda)$ through a medium of refractive index $n$ for a distance $z$ is given by

\begin{equation}\label{fourieroptics}
\tilde{E}_1(k_x, \lambda) = \tilde{E}_0(k_x, \lambda) \exp\left( \text{i} z \sqrt{\left(\frac{2\pi}{\lambda} n\right) ^2 - k_x^2} \right).
\end{equation}

The above equation allows us to propagate an optical field in a computationally inexpensive, one-step manner, which is essential due to the large distances compared to the wavelength involved in the solar spectral splitter. The theory is described further in Chapter 4 of \cite{fourier_optics}. 

When passing through a lens or an optical element, the incident field $E_{0}(x, \lambda)$ at position $x$ with wavelength $\lambda$ undergoes a phase shift. Within the thin element approximation, up to an unimportant global factor, this phase shift is given by

\begin{equation}\label{phaseshift}
E_{1}(x, \lambda) = E_{0}(x, \lambda) \exp\left( \text{i} \frac{2\pi}{\lambda} (n - 1) d(x) \right),
\end{equation}

\noindent where $d(x)$ is the thickness of the material with refractive index $n$ (See section 2.4 in \cite{fourier_optics} for details). The thin element approximation is valid when both the maximum thickness of the element $d_0$, and the angle of incidence of the light $\theta$, are sufficiently small such that $(d_0 / \lambda) \theta^2 / 2n \ll 1$. With the parameters we use, our results are robust up to angles of $\pm 20^\circ$. We furthermore ignored reflections at the material interfaces, but this is a fairly minor approximation because of the low refractive index of \textrm{SiO$_{2}$} and and the absence of very oblique angles of incidence. We verified the validity of these approximations by comparing  with full-wave simulations (see Supplementary Information).

\subsection{Solar cell current} \label{section: solar cell current}
We aim to maximize the short circuit current $J_\mathrm{sc, tot}$ of the lateral tandem setup over all considered wavelengths. We influence this current by changing the E-field $E(x, \lambda)$ reaching the lateral subcells after passing through the spectral splitter. Therefore, we need to find the theoretical description linking $E(x, \lambda)$ to $J_\mathrm{sc,tot}$. In the following we use a mathematical description, based on the explanations found in Part 2 of \cite{Solar_energy}.

The input irradiance on the top layer of the solar spectral splitter is given by the typical AM 1.5 spectrum of the sun \cite{Solar_energy} and we multiply it by a factor $\sin(\theta)$ to account for the reduction in power due to the tilt $\theta$ of the device. The considered setup through which the light propagates is shown in Fig.~\ref{fig:figure1}. As a reminder, the left lateral subcell 1 converts light in the range from $310$ to $780$ \si{\nano\meter} and the right lateral subcell 2 from $780$ \si{\nano\meter} to $1340$ \si{\nano\meter}. The black area between the two solar cells is the `dead zone', where no light is converted into current. To implement this in our code we simply multiply the field by a boxcar function representing the region where absorption occurs.

If we consider one wavelength, a light intensity $| E(x, \lambda)|^2$ arrives on the $i$th solar cell. We then integrate this over the $x$ and $y$ dimensions of the solar cell. The integration along the $y$ direction simply results in a constant factor as the device is invariant along this axis. As such, the radiant flux $\text{P}_\mathrm{i}(\lambda)$ for the $i$th solar cell is

\begin{equation}\label{I_to_E}
    \text{P}_\mathrm{i}(\lambda) = \frac{c n \varepsilon_0}{2} \iint_{x_{i,\text{L}}}^{x_{i,\text{R}}} \left| E(x, \lambda) \right|^2  \,dx\,dy,
\end{equation}

\noindent where $x_{i,\text{L}}$ and $x_{i,\text{R}}$ are the left and right boundaries of the $i$th subcell, $n$ is the refractive index, which is modeled as $1.45$ for \textrm{SiO$_{2}$}, and $\varepsilon_{0}$ is the vacuum permittivity. 
Next, we integrate $\text{P}_\mathrm{i}(\lambda)$ over all considered wavelengths and we calculate the $i$th single-junction short circuit current as

\begin{equation}\label{J_sc_i}
J_\mathrm{sc,i} = \frac{e}{h c} \int  \mathrm{EQE}(\lambda) \cdot  \lambda \cdot \text{P}_\mathrm{i}(\lambda) \,d\lambda,
\end{equation}

\noindent where $\mathrm{EQE(\lambda)}$ is the external quantum efficiency, $e$ is the elementary charge, and $h$ is the Planck constant. $\mathrm{EQE(\lambda)}$ values of up to $90\%$ are typically reached for state of the art organic photovoltaic blends. We assume a wavelength-independent performance and thus set $\mathrm{EQE(\lambda)}=0.9$.

Typically lateral multi-junction solar cells are connected in series.  Kirchhoff's current law thus enforces equal currents through both solar cells, resulting in a reverse bias to the subcell with higher $J_\mathrm{sc,i}$ when the tandem is short circuited, resulting in a reduced overall current (and reduced power conversion efficiency). As a first order approximation this effect can be modeled as $J_\mathrm{sc,tot} = \mathrm{min}_\mathrm{i}(J_\mathrm{sc,i})$. To avoid the discontinuous derivative of this model, we introduce a smooth minimum as

\begin{equation}\label{J_tot}
J_\mathrm{{sc,tot}} = J_{1} + J_{2} - \frac{1}{\alpha} \cdot \log(e^{\alpha \cdot J_{1}} + e^{\alpha \cdot J_{2}})\,.
\end{equation}

\noindent where we set the positive parameter $\alpha$ equal to $40$. This equation replicates the effect of a finite equivalent parallel conductance of the forward biased subcell. With Eq.~\ref{J_tot} we have obtained a differentiable expression for the short circuit current of the full tandem. This enables us to perform gradient based optimization to maximize $J_\mathrm{sc, tot}$.
 
\subsection{Simulation and optimization}\label{section: simulation}
Our simulation is written in Python. We use a one-dimensional simulation grid along the $x$-axis, with a width of \qty{400}{\micro\meter} and $1000$ sampling points. The two solar cells have a width of \qty{195}{\micro\meter} and a deadzone of \qty{5}{\micro\meter} in-between each cell. Notice that our simulation integrates periodic boundary conditions along the $x$-axis, meaning that a small bandgap device lies between two large bandgap devices and vice-versa. The thickness of the \textrm{SiO$_{2}$} glass between the two optimized layers is set to \qty{200}{\micro\meter}, and the distance of air between solar cells and the phase masks is \qty{2000}{\micro\meter}. See the setup in Fig.~\ref{fig:figure1}. Our simulation code and gradient calculations allow us to optimize over $50$ wavelengths and incident angles between $\pm20^{\circ}$ on a timescale of minutes.

We seek an optimal phase mask profile through gradient-based local optimization. This means we want to maximize our desired figure of merit, namely, the total current of the setup given by Eq.~\ref{J_tot}, while respecting the imposed constraints of our problem, in this case, the band gap requirements \cite{inverse_design, inverse_design2}. To optimize with any chosen optimizer, we need a method to calculate the gradients of the fields from our simulation concerning their design variables - in our case, the height profile of the phase mask. To accomplish this, we use the software package JAX \cite{jax2018github}, which automatically obtains the derivatives of native Python and NumPy code with a straightforward interface.

We use the Adam optimizer as implemented in the optax JAX package \cite{optax_deepmind2020jax} for optimization. Adam is a variant of stochastic gradient descent (SGD) that leverages momentum and adaptive learning rates for each parameter, which often outperforms naive SGD in practice \cite{adam}. For our optimization we set the learning rate \textrm{$\alpha_{t}$} to $0.1$ and kept the other parameters as given in the standard settings of the optimizer, like $\beta_{1} = 0.9$ and $\beta_{2} = 0.999$. 

\section{Author contributions}

C.R., C.J.B., L.L. and M.L.S. developed the concept of the research work. T.J.S. and M.N. developed the idea to use Fourier optics and the thin element approximation. M.L.S. planned and carried out the computations with support from T.J.S. and J.D.F. It was the initial suggestion by J.D.F. to investigate a two-layered structure. M.L.S., T.J.S., M.N. and J.D.F. verified the correctness of approximations and theory. The work was supervised by T.J.S. and C.R. M.L.S. and T.J.S. wrote the manuscript with input from C.R., M.N., J.D.F. and L.L. All authors approved the final manuscript.

\section{Acknowledgements}
The Airmass irradiance dataset was obtained from the U.S. Department of Energy (DOE)/NREL/AL-LIANCE. M.L.S. acknowledges the Karlsruhe School of Optics and Photonics (KSOP). Part of this work has been supported by the Helmholtz Association in the framework of the innovation platform “Solar TAP”. T.J.S. acknowledges funding from the Alexander von Humboldt Foundation. M.N. and C.R. acknowledge support by the Karlsruhe Institute of Technology through the “Virtual Materials Design” (VIRTMAT) project. C.J.B. and L.L. acknowledge financial support from the DFG (BR 4031/22-1 and BR 4031/21-1).

\section{Competing interests}
All authors declare no financial or non-financial competing interests. 

\section{Data availability}
All datasets  and source codes generated during the current study are available from the corresponding author upon reasonable request.

\section{Code availability}
All source code is available from the corresponding author upon reasonable request.

\printbibliography 


\end{document}


\maketitle

\textcolor{black}{\section{Data for optimized designs}
In Section~2 of our manuscript, we show the results of our optimization. The final solar spectral splitter we designed consists of two phase masks placed sequentially along the optical path, implemented by free-form structured surfaces at the top and bottom layer of a \qty{200}{\micro\meter} wide \textrm{SiO$_{2}$} substrate. More than one phase-modulating surface is essential for achieving a wide-angle tolerance. Two tandem solar cells are placed directly underneath the spectral splitter at a distance of \qty{2000}{\micro\meter}. The resulting designs are shown in Fig.~2 of the main text. Table~\ref{tab:peaks} gives the maximum peak-to-valley heights of the phase masks for the different optimizations. }

{\color{black}\begin{table} [H]
    \centering
     \taburulecolor{black}
    \begin{tabu}{>{\color{black}}c||>{\color{black}}c|>{\color{black}}c|>{\color{black}}c|>{\color{black}}c}
         \textcolor{black}{\textbf{Maximum peak-to-valley height}} & \textbf{Device 1} & \textbf{Device 2} & \textbf{Device 3} & \textbf{Device 4} \\ \hline
         Layer 1 & \qty{3.24}{\micro\meter} & \qty{4.62}{\micro\meter} & \qty{2.74}{\micro\meter} & \qty{8.47}{\micro\meter}\\
         Layer 2 & - & - & \qty{4.73}{\micro\meter} & \qty{6.23}{\micro\meter}\\
    \end{tabu}
    \captionsetup{labelfont={color=black},font={color=black}}
    \caption{The maximum peak-to-valley height is the difference between the maximal and minimal height of a surface structure along the z axis. We compare the quantity for different optimized devices. The first row ``Layer~1" shows the maximum peak-to-valley height for the top layer of the structured surfaces. The second row ``Layer~2" shows the maximum peak-to-valley height for the bottom layer of the structured surfaces. Device~1 is a one-sided phase mask optimized for $0^{\circ}$ only. Device~2 is a one-sided phase mask optimized for 50 angles between $-5^{\circ}$ and $5^{\circ}$. Since Devices 1 and 2 only have a structured top layer, we denote their maximum peak-to-valley height as ``-" for Layer~2. Device~3 is a two-sided phase mask optimized for $0^{\circ}$ only. Device~4 is a two-sided phase mask optimized for 50 angles between $-5^{\circ}$ and $5^{\circ}$. Their exact designs and performances are shown in Fig.~2 of the paper. }
    \label{tab:peaks}
\end{table}}

\section{Comparison of full-wave simulation and our approach}
As described in our paper in Section 3, we make use of the thin element method to calculate the phase shift an incident field $E_{0}(x, \lambda)$ at position $x$ with wavelength $\lambda$ undergoes when passing through a lens or an optical element. The phase shift is given by

\begin{equation}\label{phaseshift}
E_{1}(x, \lambda) = E_{0}(x, \lambda) \exp\left( \text{i} \frac{2\pi}{\lambda} (n - 1) d(x) \right)\,,
\end{equation}

\noindent where $d(x)$ is the thickness of the material with refractive index $n$. The thin element approximation holds when the maximum thickness of the element $d_0$ and the angle of incidence of the light $\theta$ are sufficiently small such that $(d_0 / \lambda) \theta^2 / 2n \ll 1$. With the parameters we use, according to this limit, our results are robust up to angles of $\pm 20^\circ$. Here, we check this theoretical prediction by comparing the phase shift from the thin element approximation to full-wave simulations using the Comsol Multiphysics software package. 

\begin{figure} [ht]
  \centering
  \begin{minipage}[t]{.6\linewidth}
      {\includegraphics[width=1.0\linewidth]{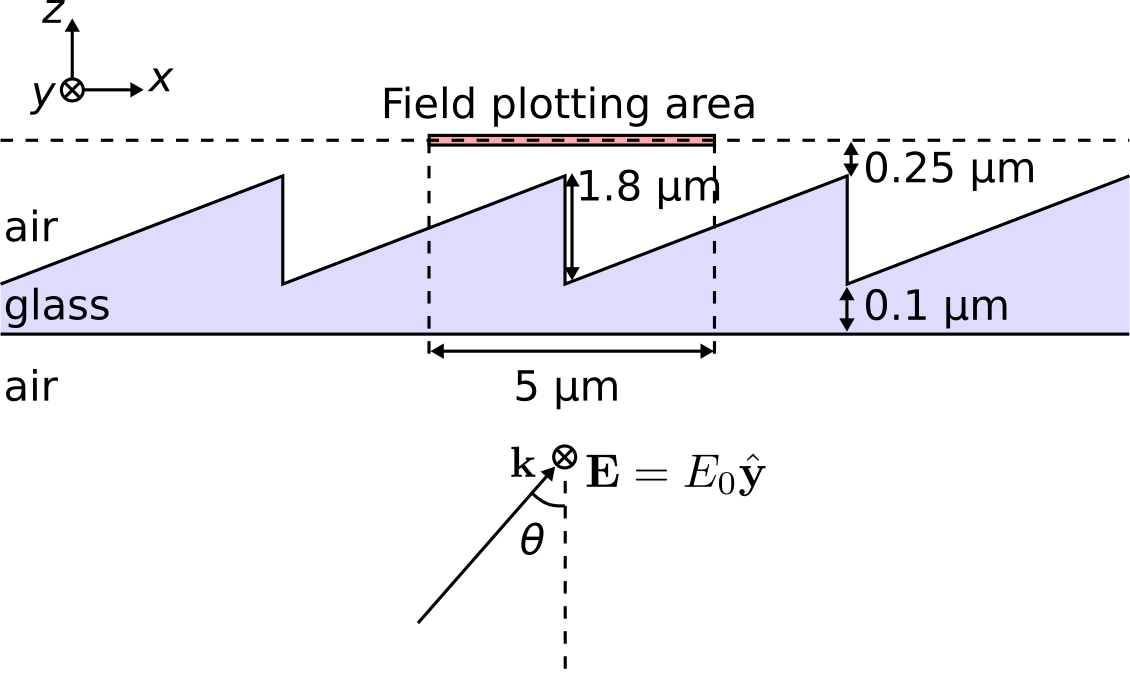}}
  \end{minipage} 
  \hfill
  \caption
    {Schematic of the considered design element.
    }%
  \label{fig:figure0}%
\end{figure}

To verify the correctness of the thin element approximation, we first selected a typical feature from the structure of the solar spectral splitter: a blazed grating-like structure with \SI{5}{\micro\meter} period and \SI{1.8}{\micro\meter} height. The actual structure is more complex as seen in the main text, but for simplicity, we assumed the structure to be periodic in this verification example. The chosen layout is depicted in Fig.~\ref{fig:figure0}, where the refractive index $n$ is chosen as $1.45$ for the glass \textrm{SiO$_{2}$} and $1$ for air.

\begin{figure} [ht]
  \centering
  \begin{minipage}[t]{.95\linewidth}
      {\includegraphics[width=1.0\linewidth]{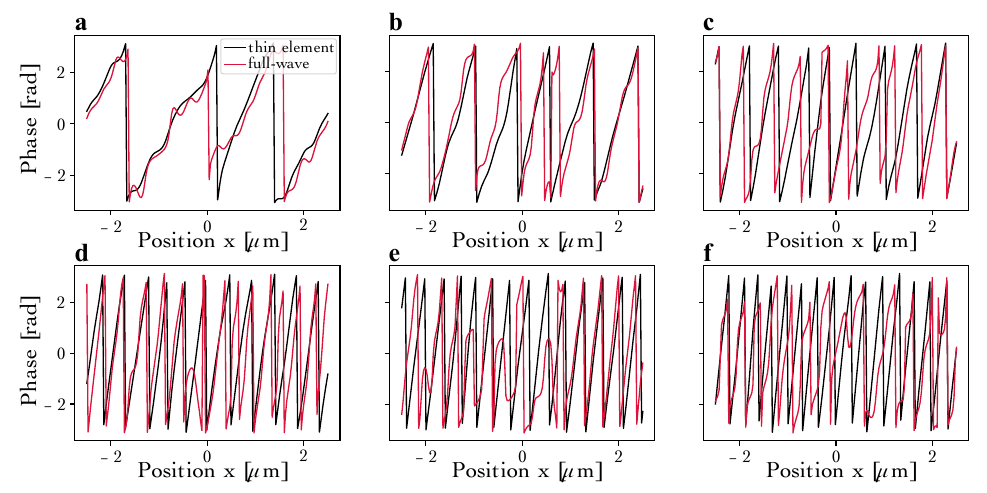}}%
  \end{minipage} 
  \hfill
  \caption
    {Comparison of thin element approximation and full-wave simulation. The top row shows the results, where approximation and full wave calculations still agree. The bottom row shows where the approximation gets less reliable. 
    (a) Normal incidence results at $0^{\circ}$. (b) Results for $10^{\circ}$ as the angle of incidence. (c) The phase for an angle of incidence at $20^{\circ}$, just around the incident angle where the approximation becomes less reliable. (d) The phase at an angle of incidence of $30^{\circ}$. (e) Results with an angle of incidence of $40^{\circ}$. (f) The phase at an angle of incidence of $50^{\circ}$. In (d)--(f), the phase shifts clearly show a large difference between thin element approximation and full wave simulation. }%
  \label{fig:figure1}
\end{figure}

We considered a monochromatic field $E_{0}(x, \lambda)$ with an amplitude of $1$ \si[per-mode = fraction]{\volt \per \meter} with a wavelength of $310$ \si{\nano\meter} to enter our model phase mask from the lower side. In the full-wave simulation, we used a $y$-polarized field. From there, we calculate the field propagation and investigate its values at $0.25$ \si{\micro\meter} distance from the grating. We then compare this result to the tools we used in our simulation, namely the propagation of an incoming field $E_{0}(x, \lambda)$ with the Fourier optics and phase shift in a material with the thin element approximation. The comparison of the phase shift for both calculation methods is shown in Fig~\ref{fig:figure1}.

We observe a good agreement for normal incidence of light in Fig.~\ref{fig:figure1}a. The thin element and full-wave simulation exhibit three peaks in the phase over the range of distance considered. When changing the incidence towards $10^{\circ}$ and $20^{\circ}$ in Figures ~\ref{fig:figure1}b and \ref{fig:figure1}c, we observe a shift of the phase shift. Nevertheless, the number of peaks and the rough structure of the phase profiles seem to agree. When considering Figures~\ref{fig:figure1}d to ~\ref{fig:figure1}f, we see differences in the phase between the full-wave and thin element methods. The onset of discrepancies means the approximation loses its relevance for angles larger than $\pm20^{\circ}$
